\documentclass[aps,prb,onecolumn,groupedaddress,showpacs]{revtex4-1}

\usepackage{graphicx,amssymb, amsmath}
\usepackage{datetime}

\newcommand{\dbar}[1]{\bar{\bar{#1}}}
\newcommand{\rmi}{{\rm i}}
\newcommand{\ff}{{\it ff}}
\renewcommand{\i}{{\it (i)}}
\newcommand{\xma}[1]{{\dbar{#1}_{\scriptscriptstyle\times}}}
\newcommand{\diag}{{\rm diag}}
\let\originalleft\left
\let\originalright\right
\renewcommand{\left}{\mathopen{}\mathclose\bgroup\originalleft}
\renewcommand{\right}{\aftergroup\egroup\originalright}
\newcommand{\parallelslant}{\mathbin{\scriptscriptstyle \!/\mkern-5mu/\!}}

\newcommand{\JLTP}{J. Low Temp. Phys. }
\newcommand{\PC}{Physica C }
\newcommand{\IEEEas}{IEEE Trans. Appl. Supercond. }
\newcommand{\JSNM}{J. Supercond. Nov. Magn. }
\newcommand{\JPSJ}{J. Phys. Soc. Jpn. }
\newcommand{\JJAP}{Jpn. J. Appl. Phys. }
\newcommand{\PRB}{Phys. Rev. B }
\newcommand{\PRL}{Phys. Rev. Lett. }
\newcommand{\APL}{Appl. Phys. Lett. }
\newcommand{\JAP}{J. Appl. Phys. }
\newcommand{\SUST}{Supercond. Sci. Technol. }
\newcommand{\ZPB}{Z. Phys. B }
\newcommand{\NatMat}{Nat. Mater. }
\newcommand{\NatPhys}{Nat. Phys. }
\newcommand{\JPCSSP} {J. Phys. C: Solid State Phys. }
\newcommand{\JETP}{Soviet Phys. JETP }
\newcommand{\EPL}{Europhys. Lett. }
\newcommand{\RPP}{Rep. Prog. Phys. }
\newcommand{\RMP}{Rev. Mod. Phys. }
\newcommand{\IEEEmag}{IEEE Trans. on Magnetics }
\newcommand{\PR}{Phys. Rev. }
\newcommand{\PL}{Phys. Lett. }

\begin{document}

\title{Theory of measurements of electrodynamic properties in anisotropic superconductors in tilted magnetic fields. Part I: flux flow and Campbell regimes.}

\author{N. Pompeo}%
\email{pompeo@fis.uniroma3.it}
\affiliation{Dipartimento di Fisica ``E. Amaldi'' and Unit\`a CNISM, Universit\`a Roma Tre, Via della Vasca Navale 84, 00195 Roma, Italy}%

\date{November 13th,  2012}

\begin{abstract}
The vortex dynamics of uniaxial anisotropic superconductors in magnetic fields applied with arbitrary orientation is theoretically studied. Focus is on the model for electrical transport experiments in the linear regime. 
Relevant vortex parameters, like the viscous drag, the vortex mobility and pinning constant (with point pins), together with the flux flow and Campbell resistivities, are derived in tensor form, in the very different free flux flow and pinned Campbell regimes.
The applicability to the various tensor quantities of the well-known scaling laws for the angular dependence on the field orientation is commented.
Moreover, it is shown that the experiments do not generally yield the intrinsic values of the anisotropic viscosity and pinning constant. Explicit expressions relating measured and intrinsic quantities are given. 
\end{abstract}

\pacs{74.25.fc, 74.25.Op, 74.25.Wx}

\maketitle

\section{Introduction}
\label{sec:intro}
Many superconductors of wide interest and recent discovery, such as iron-based superconductors \cite{ironsup}, MgB$_2$ \cite{mgb2} and cuprate superconductors,\cite{poole} have in common an intrinsic material anisotropy, essentially uniaxial, arising from their crystal structure.
The material anisotropy has a profound impact, among the others, on the vortex dynamics and on the related pinning phenomena. Such properties have been much studied due to their importance both for unraveling the fundamental physics of the underlying superconductor and in view of technological applications. 
As an example, recently a great deal of effort is devoted to the artificial tailoring of pinning on YBa$_2$Cu$_3$O$_{7-\delta}$ based coated conductors,\cite{coated1,coated2} mainly through the introduction of extended defects whose effect is to further enhance the anisotropic behaviour.

I focus on the model for electrical transport measurements in the linear regime in the mixed state, since this class of measurements is largely used in the study of vortex dynamics.
 
The interplay between the material anisotropy and the preferential direction introduced by the magnetic field $\vec{B}$ determines a non-straightforward relationship between the applied current density $\vec{J}$ and the corresponding electric field $\vec{E}$. Indeed, by applying with a generic orientations $\vec{J}$ and $\vec{B}$, the vortices move under the effect of the Lorentz force $\vec{J}\times\widehat{B}\Phi_0$ and induce (Faraday's law) an electric field $\vec{E}$ which is in general not parallel to $\vec{J}$ even in isotropic superconductors. In anisotropic superconductors, additionally, in general vortices do not move parallel to the Lorentz force, further reducing the possibility to obtain an electric field $\vec{E}$ parallel to the applied $\vec{J}$.

As a consequence, the measured quantities, such as flux flow and Campbell resistivities (and their vortex counterparts, the vortex viscosity and the pinning constant), depend on the angles between $\vec{B}$, $\vec{J}$ and the anisotropy axes. Hence they require a tensor representation and become not straightforwardly related to the material intrinsic properties.

Previous works addressed some aspects of the problem, such as: anisotropic flux flow in the pin-free d.c. regime;\cite{Ivlev1991,Genkin1989,HHTelectrodyn,HHTelectrodyn2,HHTtdgl,haoclem} pinning in non linear regimes in tilted fields, studied in the perspective of magnetization measurements;\cite{brandttensor, klupsch} two-dimensional anisotropic pinning with isotropic viscous drag and fixed magnetic field orientation;\cite{Shklovskij} coupling between anisotropic two-fluid currents and vortex motion, the latter described within an isotropic framework.\cite{CCaniso,coffeyJLTP}

In this work I propose a generalized treatment, centered on the force equation for the vortex motion, referring to uniaxial anisotropic superconductors in a magnetic field applied with generic orientation. 
Both the material anisotropy and pinning, the latter limited to weak random point pins only, are considered and studied in the very different free flux flow and pinned Campbell regimes, for arbitrary angles between $\vec{B}$, $\vec{J}$ and the anisotropy axis.

The objective is to provide all the quantities of interest as tensors, clearly distinguishing between intrinsic and measurable properties. Moreover, in several examples they will be cast into expressions directly exploitable in the experiments and applied to analyze experimental data. 

This work is organized as follows: in Section \ref{sec:rhoffi_tensor}, the electrodynamics model for the flux flow regime is recalled and recast in a suitable form; in Section \ref{sec:fluxflowregime} the vortex viscosity, vortex mobility and flux flow resistivity tensors are computed starting from the vortex force equation; in Section \ref{sec:campbell} the treatment is extended to the a.c. Campbell regime, yielding the various pinning tensors. Sections \ref{sec:exp_fluxflow} and \ref{sec:exp_Campbell} are devoted to experimental aspects, providing examples for the measurement of the mixed state anisotropic resistivity and of the pinning constant, respectively. 

\section{Vortex motion electrodynamics model}
\label{sec:rhoffi_tensor}

In their works,\cite{HHTelectrodyn,HHTelectrodyn2} Hao, Hu and Ting (HHT) addressed the problem concerning the d.c. flux flow resistivity in anisotropic superconductors in the mixed state. The model holds in the linear regime in a homogeneous superconductor with an uniform magnetic field applied along a generic direction, in the London limit. The model neglects pinning and assumes vortices to be straight and rigid flux lines moving in a uniform current field density. This basic model does not take into account more complex phenomena such as helical instabilities of the vortex lines,\cite{brandtone} which can occur when there are current components parallel to the applied field, or flux-line cutting effects,\cite{gonzalez} or the breaking of vortex lines into pancakes, occurring in the extreme anisotropy, layered superconductors.\cite{brandtone} Accordingly, I do not consider any electric field component $\parallel\vec{B}$.\cite{brandttensor}
An intrinsic flux flow conductivity tensor\cite{Note1} $\dbar{\sigma}_\ff^\i$ is introduced,\cite{HHTelectrodyn,HHTelectrodyn2}  relating the electric field $\vec{E}$ induced by vortices moving with velocity $\vec{v}$, which can be expressed through the Faraday's law as:\cite{EBv} 
\begin{equation}
\label{eq:Ev}
\vec{E}=\vec{B}\times\vec{v}
\end{equation}
to the current $\vec{J}_T$ which determines the flux flow dissipation, so that, within the linear response theory:
\begin{equation}
\label{eq:sigmaff_def}
  \vec{J}_T=\dbar{\sigma}_\ff^\i \vec{E}
\end{equation}
The inverse of $\dbar{\sigma}_\ff^\i$ yields the intrinsic flux flow resistivity tensor $\dbar{\rho}_\ff^\i$:
\begin{equation}
\label{eq:rhoff_def}
  \vec{E}=\left(\dbar{\sigma}_\ff^{\i}\right)^{\!-1}\vec{J}_T=\dbar{\rho}_\ff^\i\vec{J}_T
\end{equation}
It is important to note that $\dbar{\sigma}_\ff^\i$ and $\dbar{\rho}_\ff^\i$ are \emph{intrinsic} quantities which in general are not directly measured. Indeed, in typical experimental setups the imposed current $\vec{J}$ needs not to coincide with $\vec{J}_T$ of Eq. \eqref{eq:rhoff_def}, so that the experimentally measured resistivity tensor $\dbar{\rho}_{\ff}$, defined as:
\begin{equation}
\label{eq:rhoeff_def}
\vec{E}=\dbar{\rho}_{\ff}\vec{J}
\end{equation}
will be in general different from $\dbar{\rho}_\ff^\i$. The above equation states that in general, because of the anisotropy, $\vec{E}\nparallel\vec{J}$.

An important outcome of this paper will be to obtain explicit relations between the measured ($\dbar{\rho}_{\ff}$) and the intrinsic ($\dbar{\rho}_{\ff}^\i$) flux flow resistivity tensors. 

The fact $\vec{J}\neq\vec{J}_T$ can be understood  considering that the electric field $\vec{E}$ induced by the vortex motion is by definition perpendicular to $\vec{B}$, i.e. $\vec{E}\cdot\vec{B}=0$, and therefore:\cite{HHTelectrodyn,HHTelectrodyn2}
\begin{equation}
\label{eq:JT_def1}
\dbar{\rho}_\ff^\i\vec{J}_T\cdot\vec{B}=0
\end{equation}
The latter expression imposes that $\vec{J}_T$ cannot be freely oriented with respect to $\vec{B}$, so that $\vec{J}_T$ is in general distinct from $\vec{J}$, which instead can be externally applied with arbitrary orientation. One can write:\cite{HHTelectrodyn,HHTelectrodyn2}
\begin{equation}
\label{eq:JT_def2}
\vec{J}=\vec{J}_T+\vec{J}_S
\end{equation}
\noindent where $\vec{J}_S$ is a supercurrent density, parallel to $\vec{B}$ (i.e. $\vec{J}_S=J_S\widehat{B}$, being $\widehat{B}$ the $\vec{B}$ unit vector), giving rise to no dissipation related to vortex motion (flux line cutting\cite{gonzalez} is neglected).

\begin{figure}[ht]
\centerline{\includegraphics[width=4cm]{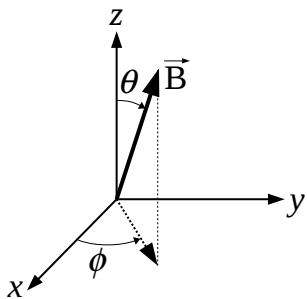}}
  \caption{ Principal frame of reference. The magnetic induction field $\vec{B}$ is also depicted, applied along a generic direction at the polar $\theta$ and azimuthal $\phi$ angles.}
\label{fig:ref}
\end{figure}

The explicit expression for the $\dbar{\sigma}_\ff^\i$ has been previously derived.\cite{Ivlev1991,Genkin1989,HHTtdgl,haoclem} I follow Ref. \onlinecite{HHTtdgl} which, working within the Time-Dependent Ginzburg-Landau theory (${B}\lesssim {B}_{c2}$), accounts for both ohmic losses and order parameter relaxation.
I first specify the frame of reference. The crystallographic axes are taken as the coordinate axes of a cartesian frame of reference (Fig. \ref{fig:ref}), so that $x\equiv a$, $y\equiv b$ and $z\equiv c$, being the latter the axis of the uniaxial anisotropy. 
In this frame of reference the phenomenological mass tensor,\cite{kogan,klemm} which can be used to describe in the London limit the material anisotropy, is diagonal:
\begin{equation}
\label{eq:masstensor}
\begin{pmatrix}
  m_{ab} & 0 & 0 \\
  0 & m_{ab} & 0 \\
  0 & 0 & m_c
\end{pmatrix}
=
m_{ab}
\begin{pmatrix}
  1 & 0 & 0 \\
  0 & 1 & 0 \\
  0 & 0 & \gamma^2
\end{pmatrix}
=
m\dbar{M}
\end{equation}
having defined the in-plane mass $m_{ab}=m$, the out-of-plane mass $m_c$ and the anisotropy factor $\gamma^2=m_c/m_{ab}$.

The mass tensor $\dbar{M}$ contains all the information concerning the material anisotropy, since the only source of anisotropy that will be considered is the effective mass of the charge carriers (this implies, for example, that the possible anisotropy of the scattering time of the normal carriers is neglected).
By neglecting the Hall contribution and assuming the same anisotropy axes for the normal resistivity and mass tensors $\dbar{\rho}_n=\rho_{n,11}\dbar{M}$ (i.e. the mass tensor is the only source of anisotropy also for the normal state), the tensors $\dbar{\rho}_\ff^\i$ and $\dbar{\sigma}_\ff^\i$ are also diagonal in the stated frame of reference. Therefore one has:
\begin{equation}
\label{eq:rhoff_tensor1}
\dbar{\rho}_\ff^\i(B, \theta)=
\begin{pmatrix}
  \rho_{\ff,11}^\i(B, \theta) & 0 & 0 \\
  0 & \rho_{\ff,11}^\i(B, \theta) & 0 \\
  0 & 0 & \rho_{\ff,33}^\i(B, \theta)
\end{pmatrix}
\end{equation}
Neglecting the weak field dependence of $\dbar{\rho}_n$ and writing down explicitly all the field dependencies, one obtains:\cite{HHTelectrodyn,HHTtdgl} 
\begin{equation}
\label{eq:rhoff_element}
\rho_{\ff,ii}^\i(B,\theta)/\rho_{n,ii}=\mathcal{F}(B/B_{c2}(\theta))
\end{equation}
\noindent 
This equation implies two relevant results. First, it is consistent with the important scaling law\cite{klemm,scaling2,BGL,blatterone} according to which the field dependence of the anisotropic physical quantities in the mixed state depends on the $B/B_{c2}(\theta)$ ratio only.
Second, it shows that all the three tensor elements of $\dbar{\rho}_\ff^\i$ share the same field dependence through a common function $\mathcal{F}(B/B_{c2})$. 

Therefore Eq. \eqref{eq:rhoff_tensor1} can be rewritten as:
\begin{eqnarray}
\label{eq:rhoff_tensor2}
\dbar{\rho}_{\ff}^\i(B,\theta)&=&\rho_{n,11}\mathcal{F}(B/B_{c2}(\theta))\dbar{M}= \nonumber \\
&=&\rho_{\ff,11}^\i(B/B_{c2}(\theta))\dbar{M}
\end{eqnarray}
The above Equation highlights the important property that the field and angular dependence of the whole intrinsic flux flow resistivity tensor can be represented by a single scalar function, namely the element $\rho_{\ff,11}^\i(B/B_{c2}(\theta))$.
Similarly, the conductivity tensor is:
\begin{equation}
\label{eq:sigma_tensor2}
\dbar{\sigma}_{\ff}^\i(B,\theta)=\sigma_{\ff,11}^\i(B/B_{c2}(\theta))\dbar{M}^{-1}
\end{equation}

In the following, for ease of notation the angular and field magnitude dependence will be explicitly written only in the Equations reporting the main results.

\section{Flux flow regime}
\label{sec:fluxflowregime}
In this section the well-known vortex force Equation,\cite{GR,Golo} involving the balance of forces acting on an individual vortex, is studied in the regime of pure flux flow. 
This regime is realized by an ideal pin-free superconductor, indifferently in d.c. and a.c. regimes, whereas in real superconductors it can be obtained with d.c. currents sufficiently large to overcome the pinning forces, or with a.c. currents at high enough frequencies.
This treatment will provide a generalized formulation which allows (i) the determination of the vortex viscosity tensor $\dbar{\eta}$ (Sec. \ref{sec:viscosity}), which gives many physical insights, and the related vortex mobility $\dbar{\mu}_v$ tensor (Sec. \ref{sec:mobility}); (ii) the subsequent computation of the measurable flux flow resistivity tensor $\dbar{\rho}_\ff$ (Sec. \ref{sec:rhoff}); (iii) the basis for the extension to a.c. regimes with pinning, such as the Campbell regime treated in Sec. \ref{sec:campbell}.

In pure, stationary, flux flow regime, with no pinning and hence no creep contributions, the viscous drag force (per unit length) $\dbar{\eta}\vec{v}$ acting on the vortices moving with velocity $\vec{v}$ exactly balances the Lorentz force $\dbar{F}_L=\Phi_0 \vec{J}\times\widehat{B}$ so that the force Equation, written in vector form, is:
\begin{equation}
\label{eq:force1}
\dbar{\eta}\vec{v}=\Phi_0 \vec{J}\times\widehat{B}
\end{equation}
In the above Equation, it is apparent that $\dbar{F}_L$ and hence the viscous drag tensor $\dbar{\eta}\vec{v}$ belong to the plane perpendicular to the magnetic field. 

\subsection{Vortex viscosity tensor}
\label{sec:viscosity}

To obtain $\dbar{\eta}$ using Eq. \eqref{eq:force1}, I consider an arbitrary field $\vec{B}=B(\sin\theta\cos\phi,\sin\theta\sin\phi,\cos\theta)$ (see Fig. \ref{fig:ref}) and a generic current density $\vec{J}$. Vortices move under the Lorentz force $\Phi_0 \vec{J}\times\widehat{B}$. Using Eq. \eqref{eq:JT_def2}, $\vec{J}\times\widehat{B}=\vec{J}_T\times\widehat{B}$ since $\vec{J}_S\parallel\vec{B}$. Then, $\vec{J}_T$ can be expressed in terms of $\vec{v}$ through Eqs. \eqref{eq:Ev} and \eqref{eq:sigmaff_def}, so that, using Eq. \eqref{eq:sigma_tensor2}:
\begin{align}
\label{eq:preeta}
\nonumber
\dbar{\eta}\vec{v}&=\Phi_0 (\dbar{\sigma}_\ff^\i(\vec{B}\times\vec{v}))\times\widehat{B}=-\Phi_0 B \xma{B}\dbar{\sigma}_\ff^\i\xma{B}\vec{v}=\\ 
&=-\Phi_0 B \xma{B}\sigma_{\ff,11}^\i\dbar{M}^{-1}\xma{B}\vec{v}=\Phi_0 B \sigma_{\ff,11}^\i\mathcal{\dbar{M}}\vec{v}
\end{align}
valid $\forall\vec{v}$.

In the above, the tensor $\xma{B}$ comes from a general property of the cross product,\cite{tensors2} according to which, given the generic vector $\vec{a}$, one has $\widehat{B}\times\vec{a}=\xma{B}\vec{a}$ with $\xma{B}$ defined from $\widehat{B}$ as described in Appendix \ref{sec:crossproduct}.
Moreover, for the sake of compactness, the tensor $\mathcal{\dbar{M}}$ has been introduced:
\begin{widetext}
\begin{equation}
\label{eq:MB}
\mathcal{\dbar{M}}(\theta,\phi)=-\xma{B}\dbar{M}^{-1}\xma{B}=
\begin{pmatrix}
\cos^2\theta+\gamma^{-2}\sin^2\phi\sin^2\theta \;\;& -\gamma^{-2}\cos\phi\sin\phi\sin^2\theta & -\cos\phi\cos\theta\sin\theta \\
-\gamma^{-2}\cos\phi\sin\phi\sin^2\theta & \cos^2\theta+\gamma^{-2}\cos^2\phi\sin^2\theta & -\sin\phi\cos\theta\sin\theta \\
-\cos\phi\cos\theta\sin\theta & -\sin\phi\cos\theta\sin\theta & \sin^2\theta
\end{pmatrix}
\end{equation}
\end{widetext}
By equating the first and last member of Eq. \eqref{eq:preeta} $\forall\vec{v}$, one obtains:

\begin{align}
\label{eq:eta}
\nonumber
\dbar{\eta}(B, \theta,\phi)&=\Phi_0 B\xma{B}\dbar{\sigma}_\ff^\i\xma{B}=\\ 
&=\Phi_0 B\sigma_{\ff,11}^\i(B/B_{c2}(\theta))\mathcal{\dbar{M}}(\theta,\phi)
\end{align}
It can be noted that the viscosity tensor $\dbar{\eta}$ does not obey to the angular scaling law and that it depends also on $\phi$, even if the superconductor is uniaxially anisotropic, because of the Faraday-Lorentz contribution included in $\mathcal{\dbar{M}}(\theta,\phi)$.
Analogously to $\dbar{\rho}_\ff^\i$ and $\dbar{\sigma}_\ff^\i$, an ``intrinsic'' viscosity tensor (diagonal like $\dbar{\sigma}_\ff^\i$) can be introduced as follows:
\begin{align}
\label{eq:eta0i} \nonumber
\dbar{\eta}^\i(B,\theta)&=\Phi_0 B \dbar{\sigma}_\ff^\i(B/B_{c2}(\theta))=\\
&=\eta^\i_{11}(B/B_{c2}(\theta))\dbar{M}^{-1}
\end{align}
which obeys the angular scaling law and whose elements are $\eta^\i_{ii}=\Phi_0 B \sigma_{\ff,ii}^\i$. Equation \eqref{eq:eta} can be thus rewritten as:
\label{eq:eta0}
\begin{align}
\nonumber
\dbar{\eta}(B,\theta,\phi)&=-\xma{B}(\theta,\phi)\dbar{\eta}^\i(B/B_{c2}(\theta))\xma{B}(\theta,\phi)\\
&=\eta^\i_{11}(B/B_{c2}(\theta))\mathcal{\dbar{M}}(\theta,\phi)
\end{align}
Before computing $\dbar{\rho}_\ff$, some relevant properties of the viscosity tensor will be highlighted.

\subsubsection{Eigenvectors and rank of the viscosity tensor}
\label{sec:viscosityeigen}

The eigenvectors of the viscosity tensor $\dbar{\eta}$ are interesting since, by definition, they are specific vortex velocities $\vec{v}_e$ for which the viscous drag force is parallel to $\vec{v}_e$, $\dbar{\eta}\vec{v}_e=\eta_e\vec{v}_e$, where the scalar $\eta_e$ is the eigenvalue of $\vec{v}_e$. Hence, for these special directions the viscosity tensor behaves as a scalar, similarly to the behaviour in isotropic materials. 

The eigenvectors and eigenvalues can be computed through standard tensor algebra,\cite{tensors} yielding the results reported in Table \ref{tab:eigeneta}.

In the Table the function $\epsilon^2(\theta)$ has been introduced:
\begin{equation}
\label{eq:epsilon}
\epsilon^2(\theta)=\cos^2\theta+\gamma^{-2}\sin^2\theta
\end{equation}
which corresponds to the well-known angular-dependent anisotropy parameter\cite{blatterone, tinkham} $\epsilon^2(\theta)$, which defines, among the others, the angle dependence of the critical field $B_{c2}(\theta)=B_{c2}(0)/\epsilon(\theta)$.
\begin{table}[h]
{\small
\begin{tabular} {l|l|l}
\hline \hline
Eigenvector $\widehat{v}_{e,i}$ & $\eta_{e,i}$ & $[\widehat{J},\widehat{v}_{e}]$ \\
 & & for $\phi=\frac{\pi}{2}$\\
\hline
1:$(\sin\phi,-\cos\phi,0)={(\widehat{B}\times\widehat{z})}/{|\widehat{B}\times\widehat{z}|}$  & $\eta_{11}^\i\epsilon^2$ & $[\widehat{B}\times\widehat{x}, \widehat{x}]$\\
2:$(\cos\theta\cos\phi,\cos\theta\sin\phi,-\!\sin\theta)\!=\!\widehat{B}\!\times\!\widehat{v}_{e1}$ & $\eta_{11}^\i$ & $[\widehat{x},\widehat{B}\times\widehat{x}]$\\
3:$(\sin\theta\cos\phi,\sin\theta\sin\phi,\cos\theta)=\widehat B$ & $0$ & N.A.\\
\hline \hline
\end{tabular}
}
\caption{First two columns: eigenvectors and corresponding eigenvalues of the viscosity tensor within uniaxial anisotropy; third column: eigenvectors computed for $\phi=\pi/2$ and corresponding current direction forming a right-handed orthogonal basis with $\widehat{B}$ and $\widehat{v}_{e}$ (see text).}
\label{tab:eigeneta}
\end{table}
It can be seen that the trivial $\widehat{v}_{e3}\parallel\widehat{B}$ has zero eigenvalue (as expected from expression \eqref{eq:MB}) and that the other two are both $\perp\widehat{B}$. Hence the three (unit) eigenvectors of Table \ref{tab:eigeneta} constitute an orthonormal right-handed basis ``d'', i.e. they are the principal axes for the viscosity tensor.
The viscosity tensor $\dbar{\eta}$ is then by definition diagonal in the frame of reference with basis ``d''.

The transformation matrix for this new frame of reference is $\dbar D=[\widehat{v}_{e1}\;\widehat{v}_{e2}\;\widehat{v}_{e3}]$,\cite{tensors} i.e. its columns are given by the coordinates (in the standard frame of Fig. \ref{fig:ref}) of the three unit eigenvectors.
In the following, the quantities represented in this frame of reference will be denoted with the superscript ``(d)'': for example, given a generic vector $\vec{a}$ and matrix $\dbar{A}$, $\vec{a}^{(d)}=\dbar{D}^{-1}\vec{a}$ and $\dbar{A}^{(d)}=\dbar{D}^{-1}\dbar{A}\dbar{D}$.\cite{tensors}
One obtains: 

\begin{align}
\label{eq:roteta}
\nonumber
{\dbar{\eta}}^{(d)}(B,\theta)&=\eta^\i_{11}(B/B_{c2}(\theta))
\begin{pmatrix}
\epsilon^2(\theta) & 0 & 0 \\
0 & 1 & 0 \\
0 & 0 & 0 
\end{pmatrix}=\\
&=\eta^\i_{11}(B/B_{c2}(\theta))\mathcal{\dbar{M}}^{(d)}(\theta)
\end{align}
It can be seen that $\dbar{\eta}^{(d)}$ is a symmetric non diagonal matrix, with zero elements in the third row and third column. The zeroes in the third row ensures that the viscous drag force component $\parallel\vec{B}$ is zero, as anticipated in the comment of Eq. \eqref{eq:force1}. The zeroes in the third column makes the viscous drag force independent from a possible component of $\vec{v}$ parallel to $\vec{B}$. Thus, the force Eq. \eqref{eq:force1} with the results \eqref{eq:eta} and \eqref{eq:roteta} imply that $\vec{v}\perp\vec{B}$, consistently with the framework of the present model in which a motion of rigid vortices parallel to their axes is meaningless. 

Moreover, it is evident that the viscosity tensor, defined within the three-dimensional space, is not invertible, since its rank=2. Hence the viscosity tensor is a bijection not in the whole three-dimensional space but only within the plane perpendicular to $\widehat{B}$. This implies that, for a given Lorentz force, the corresponding vortex velocity is univocally determined, property which is useful in the computation of the vortex mobility tensor.

Going back to the eigenvectors, since $\widehat{v}_{e1}$ and $\widehat{v}_{e2}$ are  $\perp\widehat{B}$, the force Equation \eqref{eq:force1} implies that for both of them it is possible to find a proper current direction $\widehat{J}$ which gives a right-handed orthogonal basis $[\widehat{J},\widehat{B},\widehat{v}_{e}]$, which evidently ensures that $\vec{E}\parallel\vec{J}$. 

\begin{figure}[ht]
\centerline{\includegraphics[width=8cm]{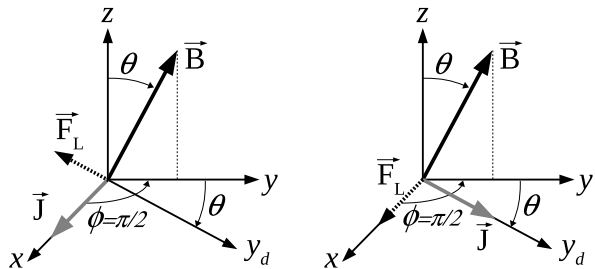}}
  \caption{ Specific field-current configurations (see text).}
\label{fig:fieldcurconf}
\end{figure}

Taking $\vec{B}$ in the $y$-$z$ plane ($\phi=\pi/2$) for the ease of notation and without loss of generality, these current directions are reported in the third column of Table \ref{tab:eigeneta} and represented in Fig. \ref{fig:fieldcurconf}, where the left and right panels corresponds to the second line and first line in the Table, respectively. In the Figure, $\widehat{y}_d=(\widehat{B}\times\widehat{x})\parallel\widehat{J}$ has been introduced for later reference.
It is worth stressing again that in general $\vec{E}\nparallel\vec{J}$, which makes more complex the experimental determination of the material resistivity. In this view, the above two special current-field configurations are especially useful.

\subsubsection{Vortex viscosity along the principal axes}
\label{sec:vortexprincipal}
I now consider specific current-field configurations where both vectors are directed along the principal axes of a uniaxial superconductor. I compare the viscosity obtained from the tensor analysis to the results\cite{haoclem} computed on the bases of the Bardeen\textendash{}Stephen\cite{BS} model. From Table \ref{tab:eigeneta} and Eq. \eqref{eq:eta0}, it is easy to check that for these configurations one has the usual right-handed orthogonal basis $[\widehat{J},\widehat{B},\widehat{v}]$ and that $\dbar{\eta}\vec{v}=\eta_e\vec{v}$. As it can be seen in Table \ref{tab:etacfr}, the results obtained through the two above mentioned procedures are coherent. In the Table the relation $\eta_{11}^\i(\frac{\pi}{2})=\eta_{11}^\i(0)\gamma$, easily obtainable within the Bardeen\textendash{}Stephen model, is used.

One additional comment can be done by considering the full $\dbar{\eta}$ tensor explicitly computed for the field orientations along the principal axes and reported in the fifth column of Table \ref{tab:etacfr}. In this case, the viscosity tensor is diagonal: this means that, by taking $\widehat{B}$ and $\widehat{J}$ along two principal axes, the vortex motion (both in terms of $\widehat{v}$ and $\widehat{F}_L$) occurs along the third principal axis.

Moreover, when $\widehat{B}$ is parallel to a principal axis, it can be easily verified that the viscosity tensor $\dbar{\eta}$ is diagonal if \emph{and only if} $\dbar{\sigma}_\ff^\i$ is itself diagonal, meaning that the same principal axes of $\dbar{\sigma}_\ff^\i$ are inherited by $\dbar{\eta}$. This property will be exploited when treating the pinning constant tensor in Sec. \ref{sec:campbell}.
\begin{table*}
\begin{tabular} {c|c|l|l|c}
\hline \hline
Set & angles & $\eta$ from Ref. \onlinecite{haoclem}& $\eta_e$ from Table \ref{tab:eigeneta}& $\dbar{\eta}$ \\
$[\widehat{J},\widehat{B},\widehat{v}]$ & $\theta,\phi$ & &  and Eq. \eqref{eq:eta0} & \\
\hline
x,z,y & 0,.. &$\eta_b^{(c)}$ & $\eta_{11}^\i(0) $ & $\diag(\eta_{11}^\i(0),\eta_{11}^\i(0), 0)$ \\
x,y,z & $\frac{\pi}{2},\frac{\pi}{2}$ & $\eta_c^{(b)}=\eta_c^{(a)}\approx\eta_b^{(c)}\gamma$ & $\eta_{11}^\i(\frac{\pi}{2})=\eta_{11}^\i(0)\gamma$ & $\diag(\eta_{33}^\i(\frac{\pi}{2}), 0, \eta_{11}^\i(\frac{\pi}{2}))$\\
z,y,x & $\frac{\pi}{2},\frac{\pi}{2}$ & $\eta_a^{(b)}\approx\eta_b^{(c)}\gamma^{-1}$ & $\eta_{33}^\i(\frac{\pi}{2})=\eta_{11}^\i(0)\gamma^{-1}$ & $\diag(\eta_{33}^\i(\frac{\pi}{2}), 0, \eta_{11}^\i(\frac{\pi}{2}))$\\
z,x,y & $\frac{\pi}{2},0$ & $\eta_b^{(a)}\approx\eta_b^{(c)}\gamma^{-1}$ & $\eta_{33}^\i(\frac{\pi}{2})=\eta_{11}^\i(0)\gamma^{-1}$ & $\diag(0, \eta_{33}^\i(\frac{\pi}{2}), \eta_{11}^\i(\frac{\pi}{2}))$\\
\hline \hline
\end{tabular}
\caption{First two columns: geometric configuration; third column: $\eta$ from Ref. \onlinecite{haoclem}, following the notation used there (i.e. subscript and superscript denote velocity and field directions, respectively); fourth column: the scalar viscosity $\eta_e$ obtained from Table \ref{tab:eigeneta} and Eq. \eqref{eq:eta0}; fifth column: full $\dbar{\eta}$ tensor.}
\label{tab:etacfr}
\end{table*}

\subsection{Vortex mobility tensor}
\label{sec:mobility}
The force Equation \eqref{eq:force1} can be written in terms of the vortex mobility tensor $\dbar{\mu}_v$ as:
\begin{equation}
\label{eq:force2}
\vec{v}=\dbar{\mu}_v\Phi_0 \vec{J}\times\widehat{B}
\end{equation}
Since the vortex viscosity tensor has rank=2, i.e. it is not invertible, $\dbar{\mu}_v$ cannot straightforwardly be derived as the inverse of $\dbar{\eta}$, as the relation between the corresponding scalar quantities would suggest.
Thus, I exploit the results of Section \ref{sec:viscosity}: within the plane $\perp B$, $\dbar{\mu}_v$ is the inverse bijection of $\dbar{\eta}$ whereas 
the vortex mobility for the (physically not existing) Lorentz force component $\parallel\widehat{B}$ can be safely set $=0$. Hence, in the frame of reference ``(d)'' the vortex mobility is the following diagonal matrix:
\begin{equation}
\label{eq:mu_d}
\dbar{\mu}_{v}^{(d)}=\left(\eta^\i_{11}\right)^{\!-1}\diag(\epsilon^{-2},1,0)
\end{equation}
In the principal frame of reference, one obtains:
\begin{equation}
\label{eq:mu_expr}
\dbar{\mu}_{v}=\dbar{D}\dbar{\mu}_{v}^{(d)}\dbar{D}^{-1}
\end{equation}

\subsection{Measured flux flow resistivity tensor}
\label{sec:rhoff}
The flux flow resistivity tensor $\dbar{\rho}_\ff$ can be now computed in terms of the vortex mobility $\dbar{\mu}_{v}$. By left cross-multiplying both members of Eq. \eqref{eq:force2} by $\vec{B}$ one obtains, after a little algebra, $\vec{E}=-\xma{B}\dbar{\mu}_v\Phi_0B\xma{B}\vec{J}$, valid $\forall\;\vec{J}$, which implies:
\begin{equation}
 \label{eq:rhopff2}
\dbar{\rho}_\ff=-\xma{B}\dbar{\mu}_v\Phi_0B\xma{B}
\end{equation}
It is of paramount importance that Eq. \eqref{eq:rhopff2} represents the \emph{measured}, \emph{apparent}, flux flow resistivity tensor, \emph{different} from the intrinsic flux flow resistivity tensor. The relation between $\dbar{\rho}_\ff$ (measured) and $\dbar{\rho}_\ff^\i$ (intrinsic) can be made explicit as follows. In the frame of reference ``d'', considering that $\dbar{D}^{-1}\xma{B}\dbar{D}=\xma{z}$ and using Eq. \eqref{eq:mu_d}, one obtains:
\begin{align*}
\label{eq:rhopff2_1}
\dbar{\rho}_\ff&=\Phi_0 B \dbar{D}\left(-\xma{z}\dbar{\mu}_{v}^{(d)}\xma{z}\right)\dbar{D}^{-1}=\\
&=\Phi_0 B \dbar{D}\frac{\diag(1,\epsilon^{-2},0)}{\eta^\i_{11}}\dbar{D}^{-1}=\\
&=\Phi_0 B \frac{\dbar{D}\dbar{\eta}^{(d)}\dbar{D}^{-1}}{\left[\eta_{11}^\i\epsilon\right]^2}= \frac{\Phi_0B\dbar{\eta}}{\left[\eta_{11}^\i\epsilon\right]^2}
\end{align*}
Using Eqs. \eqref{eq:eta0i} and \eqref{eq:eta0} the role of $\dbar{\sigma}_\ff^{(i)}$ emerges:
\begin{equation}
\label{eq:rhopff2_2}
\dbar{\rho}_\ff=\left(\frac{\Phi_0 B}{\eta_{11}^\i}\right)^{\!2} \frac{-\xma{B}\dbar{\sigma}_\ff^\i\xma{B}}{\epsilon^2}
\end{equation}
so that, in terms of $\dbar{\rho}_\ff^{(i)}=\left(\dbar{\sigma}_\ff^\i\right)^{\!-1}$:
\begin{equation}
\label{eq:rhoff_uni}
\dbar{\rho}_{\ff}(B,\theta,\phi)=\rho_{\ff,11}^\i(B/B_{c2}(\theta))\frac{\dbar{\mathcal{M}}(\theta,\phi)}{\epsilon^2(\theta)}
\end{equation}
It can be noted that $\dbar{\rho}_{\ff}$, like the viscosity tensor $\dbar{\eta}$ of Eq. \eqref{eq:eta}, does not obey to the angular scaling law.
This expression for the measurable flux flow resistivity tensor is particularly useful and as such will  be largely used and extended in the rest of this work. It can be checked that, as shown in Appendix \ref{sec:HHT_rhoff}, the same result can be obtained following the whole HHT model but, as already mentioned, the present derivation from the force Equation makes possible extensions to a.c. regimes as it will be done in Sec. \ref{sec:campbell}.

It is worthwhile to stress that Eq. \eqref{eq:rhoff_uni} implies that the measured flux flow resistivity tensor $\dbar{\rho}_{\ff}$ is symmetric (respecting the Onsager principle \cite{onsagercit}) but non-diagonal even if one has chosen to neglect the Hall terms in the intrinsic (diagonal) tensor $\dbar{\rho}_{\ff}^\i$. This important property describes the physical phenomenon that, by applying with generic orientations $\vec{J}$ and $\vec{B}$ to a superconductor (either isotropic or anisotropic), the vortex motion induces an electric field $\vec{E}$ which is  in general not parallel to $\vec{J}$. Moreover, in anisotropic superconductors vortices do not move in general parallel to the Lorentz force direction $\widehat{J}\times\widehat{B}$, further reducing the possibility to obtain $\vec{E}$ parallel to $\vec{J}$.

\section{Determination of the intrinsic resistivity tensor elements from experimental measured quantities}
\label{sec:exp_fluxflow}

In this section I relate the experimentally obtainable quantities to the intrinsic resistivity tensor, in various typically accessible experimental configurations. The results, Eqs. \eqref{eq:rhoeff_exp3}-\eqref{eq:rho3extr}, give the tools to extract the tensor elements $\rho_{\ff,ii}$ from the experiments. 
The issue of the true current path is of paramount importance, and it has to be solved separately. \cite{esposito} Here, I focus on the scalar resistivity $\rho^{(\widehat{J})}$ measured along the direction of the applied current $\vec{J}$, which is computed in general as:
\begin{equation}
\label{eq:rhoJ}
\rho^{(\widehat{J}) }=\vec{J}\cdot\vec{E}/J^2=\widehat{J}\cdot\vec{E}/J=\left(\dbar{\rho}\widehat{J}\right)\cdot\widehat{J}
\end{equation}
where in the flux flow regime $\dbar{\rho}=\dbar{\rho}_\ff$ (measurable).

Since $\vec{E}\perp\vec{B}$, then $\vec{J}\cdot\vec{E}=\vec{J_T}\cdot\vec{E}$, coherently with the fact that $\vec{J_T}$ is involved in the vortex motion dissipation (Eq. \eqref{eq:sigmaff_def}). In the above equation, $\widehat{J}\cdot\vec{E}$ is the electric field component, parallel to $\vec{J}$, which has to be experimentally measured in order to determine $\rho^{(\widehat{J})}$.

First I consider a static field $\vec{B}=B(\sin\theta\cos\phi,\sin\theta\sin\phi,\cos\theta)$ with generic orientation (depicted in Fig. \ref{fig:ref}) and an external current $\vec{J}=J\widehat{x}$ applied along the $x$-axis. This configuration is commonly employed to obtain ``$a$-$b$ plane'' properties.

The resistivity $\rho^{(x)}$, where the superscript ``($x$)'' refers to the orientation of the applied current, can be computed through Eq. \eqref{eq:rhoJ} with $\dbar{\rho}=\dbar{\rho}_\ff$ given by Eq. \eqref{eq:rhoff_uni}, yielding:
\begin{equation}
\label{eq:rhoeff_exp3}
  \rho^{(x)}(\theta,\phi)\!=\!\rho_{\ff,11}^\i(\theta) \frac{\rho_{\ff,11}^\i(\theta)\sin^2\!\theta\sin^2\!\phi+\rho_{\ff,33}^\i(\theta)\cos^2\!\theta}{\rho_{\ff,11}^\i(\theta) \sin^2\!\theta+\rho_{\ff,33}^\i(\theta)\cos^2\!\theta}
\end{equation}
This result, here obtained through tensor algebra, coincides with Ref. \onlinecite{HHTelectrodyn, HHTelectrodyn2}. Equation \eqref{eq:rhoeff_exp3} shows the very important result: the measured resistivity is in general a mixture of the intrinsic flux flow resistivity tensor elements $\rho_{\ff,11}^\i$ (in-plane or $a$-$b$ plane resistivity) and $\rho_{\ff,33}^\i$ (out-of-plane or $c$-axis resistivity), even if the external current flows parallel to the $a$-$b$ planes.

The intrinsic resistivity $\rho_{\ff,11}^\i(\theta)$ can be directly accessed by making measurements within the well-known ``maximum Lorentz force'', where $\vec{J}=J\widehat{x}$ as before and in addition $\vec{B}(\theta,\phi=\pi/2)=B(0,\sin\theta,\cos\theta)\in y$-$z$ plane, so that the magnetic field forms, as demonstrated in Sec. \ref{sec:viscosityeigen}, a right-handed orthogonal basis $[\widehat{J}\equiv\widehat{x},\widehat{B},\widehat{v}\equiv\widehat{F}_L]$ with the current and the vortex velocity (the latter being an eigenvector of $\dbar{\eta}$).

This configuration is represented in the left panel of Fig. \ref{fig:fieldcurconf}. One obtains:
\begin{equation}
\label{eq:rhoff_x}
  \rho^{(x)}(\theta,\pi/2)=\rho_{\ff, 11}^\i(\theta)
\end{equation}
This is the only easily accessible configuration where one has direct access to the $a$-$b$ plane intrinsic flow resistivity.

Another typical configuration is $\vec{J}\parallel\widehat{z}$,\cite{caxisrho} often used to study $\rho_{\ff,33}^\i$. The result is, independently from the angle $\phi$:
\begin{equation}
\label{eq:rhoff_z}
    \rho^{(z)}(\theta)=\frac{\rho_{\ff,11}^\i(\theta)\rho_{\ff,33}^\i(\theta) \sin^2\theta}{\rho_{\ff,11}^\i(\theta) \sin^2\theta+\rho_{\ff,33}^\i(\theta)\cos^2\theta}
\end{equation}
which yields $\rho_{\ff,33}^\i$ only for $\rho^{(z)}(\pi/2)=\rho_{33}^\i(\pi/2)$, i.e. for $\vec{B}\in x$-$y$ plane. Observing this Equation, it can be seen that it is not possible to directly measure $\rho_{\ff,33}^\i(\theta)$ for each $\theta$, contrary to what happens for $\rho_{\ff,11}^\i(\theta)$. 

Experimental results analogous to those represented by Eq. \eqref{eq:rhoff_z} can be obtained by taking  $\vec{J}=J\widehat{x}$ and $\vec{B}(\theta,\phi=0)=B(\sin\theta, 0, \cos\theta)\in x$-$z$ plane, so that the magnetic field forms the (varying) angle $\theta$ with the current $\vec{J}$:
\begin{equation}
\label{eq:rhoff_xz}
    \rho^{(x)}(\theta,0)=\frac{\rho_{\ff,11}^\i(\theta)\rho_{\ff,33}^\i(\theta) \cos^2\theta}{\rho_{\ff,11}^\i(\theta) \sin^2\theta+\rho_{\ff,33}^\i(\theta)\cos^2\theta}
\end{equation}
This is another common experimental configuration,\cite{lorentzXZ} which again yields an experimental resistivity arising from the admixture of $\rho_{\ff,11}^\i$ and $\rho_{\ff,33}^\i$.

Indeed, with a little algebra (see Appendix \ref{sec:tensorelementsderivation}) it can be demonstrated that the direct measurement of $\rho_{\ff,33}^\i(\theta)$ is impossible in general: whichever combination of directions of $\vec{B}$ and $\vec{J}$ is chosen,  $\rho_{\ff,33}^\i(\theta)$ cannot be directly accessed: the measured angular dependence of the vortex-state, $c$-axis resistivity in tilted fields \emph{never} corresponds to the material-dependent quantity.

Hence, $\rho_{\ff,33}^\i(\theta)$ must be {\em indirectly} determined through coupled, complementary measurements. An obvious choice for one of the two needed configurations is the maximum Lorentz force configuration (Eq. \eqref{eq:rhoff_x}), while the other could be equivalently chosen between the two last commented configurations (Eqs. \eqref{eq:rhoff_z} and \eqref{eq:rhoff_xz}).
Another possible choice is the configuration discussed in the study of the viscosity tensor eigenvectors (see Sec. \ref{sec:viscosityeigen}): $\vec{B}(\theta,\pi/2)$ and $\vec{J}\in y$-$z$ plane along the $\widehat{y}_d=\widehat{B}\times\widehat{x}$ direction, as depicted in the right panel of Fig. \ref{fig:fieldcurconf}, yielding the right-handed orthogonal basis $[\widehat{J}\equiv\widehat{y}_d,\widehat{B},\widehat{v}\equiv\widehat{F}_L\equiv\widehat{x}]$. Although it is seldom used in the experiments (requiring samples grown with the c-axis \emph{parallel} to the sample surface\cite{aaxis}), it will be useful in the subsequent discussion about the Campbell resistivity tensor (see Section \ref{sec:campbell}) since it ensures $\vec{E}\parallel\vec{J}$ contrary to the two configurations of Eqs. \eqref{eq:rhoff_z} and \eqref{eq:rhoff_xz}.

In this case one obtains:
\begin{equation}
\label{eq:rhoff_y}
    \rho^{(y_d)}(\theta)=\frac{\rho_{\ff,11}^\i(\theta)\rho_{\ff,33}^\i(\theta)}{\rho_{\ff,11}^\i(\theta) \sin^2\theta+\rho_{\ff,33}^\i(\theta)\cos^2\theta}
\end{equation}
By combining Eqs. \eqref{eq:rhoff_x} and \eqref{eq:rhoff_y}, the intrinsic $\rho_{\ff,33}^\i(\theta)$ can be written down in terms of the measured resistivities:
\begin{equation}
\label{eq:rho3extr}
    \rho_{\ff,33}^\i(\theta)=\frac{\rho^{(y_d)}(\theta)\rho^{(x)}(\theta)\sin^2\theta}{\rho^{(x)}(\theta)-\rho^{(y_d)}(\theta)\cos^2\theta}
\end{equation}
An analogous result can be obtained by using Eq. \eqref{eq:rhoff_z} or \eqref{eq:rhoff_xz} instead of Eq. \eqref{eq:rhoff_y}. It is also worth noting that the above Eq. \eqref{eq:rho3extr}, because of the subtraction in the denominator,\cite{Note2} yield an indeterminate form $0/0$ for $\theta=0$, implying that $\rho_{\ff,33}^\i(\theta=0)$ cannot be measured. This is true in general: it can be easily seen by writing down the measurable resistivity tensor (Eq. \eqref{eq:rhoff_uni}) for $\theta=0$, obtaining $\dbar\rho_\ff(\theta=0)=\diag(\rho_{\ff,11}^\i(0),\rho_{\ff,11}^\i(0),0)$.\cite{Note3} Since $\rho_{\ff,33}^\i(0)$ does not appear in it, no current direction can yield the $c$-axis resistivity for $\theta=0$. Hence the material resistivity along the anisotropy axis, i.e. the intrinsic $c$-axis resistivity, cannot be measured if the magnetic field is parallel to the same axis, not even by resorting to the measurement of the transverse resistivity components (i.e. through the determination of the 
transverse components of the electric field).

On the other hand, for a generic $\theta\neq0$ it has been shown that the correct process to determine the intrinsic $c$-axis resistivity $\rho_{\ff,33}^\i(\theta)$ necessarily requires the combined measurements of the resistivities with two distinct current-field setups.

\section{Campbell regime}
\label{sec:campbell}

A totally different limit with respect to flux-flow is given by the pinning regime. There, the d.c. response is zero. However, the a.c. resistivity is well measurable (Ref. \onlinecite{Golo} and references therein). To my knowledge, no complete treatise of the anisotropic resistivity in the pinning regime exists. In this Section I consider the vortex motion under a.c. currents flowing in anisotropic superconductors with random point pinning centers, described in the weak collective pinning regime.\cite{blatterone} In particular, I focus on the well-known Campbell regime, in which the pinning action on vortices dominates over the dissipative viscous drag and the pinning force is proportional to the small displacement $\vec{u}$ of the vortices from the pinning centers. The above two conditions are achieved respectively for frequencies smaller than the so-called pinning frequency (of the order of a few MHz in conventional superconductors\cite{GR} and of several GHz in high-$T_c$ superconductors,\cite{Wu1995,
Golo,
omegap_HTS,omegap_kp_HTS} low-$T_c$ thin films\cite{lowTc_omegap} and more complex superconducting heterostructures\cite{trilayers}) and for currents small enough to ensure the validity of the linear regime. Considering only point pins (i.e. zero-dimensional pinning centers) ensures that no further preferential directions are introduced in the superconducting system, in contrast with extended (e.g. linear or planar) defects.

The starting point is the force equation written in the sinusoidal regime $e^{\rmi\omega t}$, including the pinning force and neglecting the viscous drag:
\begin{equation}
\label{eq:force3}
\frac{1}{\rmi\omega}\dbar{k}_p\vec{v}=\Phi_0 \vec{J}\times\widehat{B}
\end{equation}
where $\dbar{k}_p$ is the pinning constant (also called Labusch parameter) tensor and $\vec{u}= \vec{v}/(\rmi \omega)$ is the displacement of the fluxon from the pinning center. No creep phenomena are considered.

In the Campbell regime, losses due to finite real conductivity are neglected, so that the resistivity is purely imaginary and defined, in isotropic superconductors, as:
\begin{equation}
\label{eq:rhoC}
  \rho_C=\frac{\Phi_0 B}{k_p}\omega=\omega\mu_0\lambda_C^2
\end{equation}
where $\lambda_C$ is the Campbell penetration depth.\cite{Campbellpenetration} 

Going back to anisotropic superconductors, it is evident that the force Equation \eqref{eq:force3} in the Campbell regime is formally equivalent to the force Equation \eqref{eq:force1} written for the pure flux flow regime.
Hence, by extending to the a.c. regime the electrodynamics model for the electrical transport in the mixed state extensively discussed and exploited in Sec. \ref{sec:fluxflowregime}, it is straightforward to introduce the series of tensors $[\dbar{k}_p/(\rmi\omega), \dbar{k}_p^\i/(\rmi\omega), -\rmi\dbar{\sigma}_C^\i, \rmi\dbar{\rho}_C^\i, \rmi\dbar{\rho}_C]$ 
which are dual, both in terms of roles of the tensors and of their relationships, to the already studied series $[\dbar{\eta}, \dbar{\eta}^\i, \dbar{\sigma}_\ff^\i, \dbar{\rho}_\ff^\i, \dbar{\rho}_\ff]$.
Therefore the following expressions, analogous to Eqs. \eqref{eq:eta0}, \eqref{eq:eta0i}, \eqref{eq:rhoff_def} and \eqref{eq:rhopff}, hold:

\begin{subequations}
\label{eq:pinningtensors}
\begin{align}
\label{eq:kptensor}
\dbar{k}_{p}&=-\xma{B}\dbar{k}_{p}^\i\xma{B} \\
\label{eq:kptensori}
\dbar{k}_{p}^\i&=\omega\Phi_0B\dbar{\sigma}_{C}^{\i}\\
\label{eq:rho_Ci}
\dbar{\rho}_C^\i&=\left(\dbar{\sigma}_{C}^{\i}\right)^{-1} \\
\label{eq:rho_C}
\dbar{\rho}_{C}&=-\xma{B}\left(\frac{\left|\dbar{\rho}_C^\i\right|\left(\dbar{\rho}_C^{\i}\right)^{-1}}{\left(\dbar{\rho}_C^\i\widehat{B}\right)\cdot\widehat{B}}\right)\xma{B}
\end{align}
\end{subequations}
where the latter expression holds for a diagonal $\dbar{\rho}_C^\i$, as will be taken in the following.
It is worth noting that the above Equations do not complete the model in the Campbell regime, since an explicit expression for the intrinsic tensor $\dbar{\rho}_C^\i$ is still needed. 
Whereas in the flux flow regime the problem about the corresponding intrinsic tensor $\dbar{\rho}_\ff^\i$ was addressed by the TDGL treatment,\cite{HHTtdgl} the determination of the intrinsic tensor $\dbar{\rho}_C^\i$ will be addressed in the following Section.

\subsection{The Campbell resistivity tensor elements}
\label{sec:campbellelements}

I assume that $\dbar{\rho}_C^\i$ is diagonal like $\dbar{\rho}_\ff^\i$: this ensures that, according to the final comment of Sec. \ref{sec:vortexprincipal}, choosing $\widehat{B}$ and $\widehat{J}$ along two principal axes, the corresponding pinning force (and vortex velocity) will be parallel to the third axis, as it is reasonable to expect. This assumption does not define the whole tensor, since it does not give any information about the two (within the uniaxial anisotropy) non-zero diagonal elements, $\rho_{C,11}^\i(\theta)$ and $\rho_{C,33}^\i(\theta)$.

As already seen for its flux flow counterpart, $\rho_{C,33}^\i(\theta)$ is not directly accessible through real current-field configurations. 
Therefore, it has to be derived indirectly starting from two experimentally measurable resistivities. I choose $\rho_{C}^{(x)}(\theta)$ and $\rho_{C}^{(y_d)}(\theta)$, which are defined within the two specific field-current configurations ($\phi=\pi/2$) commented in Sec. \ref{sec:exp_fluxflow} and depicted in Fig. \ref{fig:fieldcurconf}. It is worth recalling that, in the present Campbell regime, the applied $\vec{J}$ is a (low-frequency and small intensity) a.c. current.
Rewriting Eqs. \eqref{eq:rhoff_x} and \eqref{eq:rhoff_y} for $\dbar{\rho}_C$, one has:
\begin{subequations}
\label{eq:rhoC_i}
\begin{align}
\label{eq:rhoC_i1}
  \rho_{C,11}^\i(B,\theta)&=\rho_{C}^{(x)}(B,\theta) \\
  \rho_{C,33}^\i(B,\theta)&=\frac{\rho_{C}^{(x)}(B,\theta)\sin^2\theta}{\frac{\rho_{C}^{(x)}(B,\theta)}{\rho_{C}^{(y_d)}(B,\theta)}-\cos^2\theta}
\end{align}
\end{subequations}
According to Sec. \ref{sec:viscosityeigen}, both the field-current configurations ``(x)'' and ``(y$_d$)'', related to $\rho_{C}^{(x)}$ and $\rho_{C}^{(y_d)}$, correspond to eigenvectors of the vortex velocity $\vec{v}$, and therefore of the vortex displacement $\vec{u}\parallel\vec{v}$. Hence the pinning force $\dbar{k}_p\vec{u}=k_{p,e}\vec{u}$ is $\parallel \vec{u}$ and is completely described by the scalar eigenvalue pinning constant $k_{p,e}$, which within the two configurations will be denoted as  $k_p^{(x)}$ or $k_p^{(y_d)}$. This property explains the present choice of the field-current configurations, since other geometries would have given different orientations for $\vec{u}$ yielding $\dbar{k}_p\vec{u}\nparallel\vec{u}$ and thus preventing a description of the pinning force through one scalar $k_{p,e}$ only.  

The definition \eqref{eq:rhoC} allows then to write down the following:
\begin{subequations}
\label{eq:rhoC_xy}
\begin{align}
\label{eq:rhoC_xya}
  \rho_{C}^{(x)}(B,\theta)&=\frac{\Phi_0 B}{k_p^{(x)}(B,\theta)}\omega \\
  \rho_{C}^{(y_d)}(B,\theta)&=\frac{\Phi_0 B}{k_p^{(y_d)}(B,\theta)}\omega
\end{align}
\end{subequations}
The actual values of the pinning constants are determined as follows. 

\subsection{The pinning constant tensor elements}

By simple physical arguments,\cite{koshelev} for an isotropic superconductor one can evaluate the pinning constant $k_p$ by equating the maximum pinning force $k_p r_p$ acting on a vortex, where $r_{pin}$ denotes the action range of the pinning centers, with the maximum Lorentz force $J_c\Phi_0$ exerted when the current equals the critical current density. Since for core-pinning $r_{pin}\sim\xi$, i.e. the coherence length which define the length scale for the spatial variations of order parameter and therefore the radius of the vortex core, one can write
\begin{equation}
\label{eq:kp_Jc}
  k_p=c\frac{\Phi_0 J_c}{\xi}
\end{equation}
where $c\sim 1$. 

For anisotropic superconductors, Eq. \eqref{eq:kp_Jc} must be specialized for the two current configurations ``(x)'' and ``(y$_d$)'' used in Eqs. \eqref{eq:rhoC_xy}. 
The pinning constant $k_p^{(x)}$ [$k_p^{(y_d)}$] describes the pinning action on vortices moving along the $y_d$ [$x$] direction under the action of a current $\parallel x$ [ $\parallel y_d$]: hence the critical current density $J_c^{(x)}\parallel x$ [$J_c^{(y_d)}\parallel y_d$], and $r_{pin}\sim\xi^{(y_d)}$ [$r_{pin}\sim\xi^{(x)}$], i.e. the coherence length along the direction $y_d$ [$x$] of the vortex movement. Therefore:
\begin{subequations}
\label{eq:kp_Jc_xy}
\begin{align}
\label{eq:kp_Jc_xya}
  k_p^{(x)}(B,\theta)&=c\frac{\Phi_0 J_c^{(x)}(B,\theta)}{\xi^{(y_d)}(\theta)}\\
\label{eq:kp_Jc_xyb}
  k_p^{(y_d)}(B,\theta)&=c\frac{\Phi_0 J_c^{(y_d)}(B,\theta)}{\xi^{(x)}(\theta)}
\end{align}
\end{subequations}
One should note that $k_p^{(y_d)}$, despite being referred to a current-field configuration of difficult realization in the experiments, is related to quantities ($J_c^{(y_d)}$ and $\xi^{(x)}$) which can be expressed within available theories. In this sense, Eq. \eqref{eq:kp_Jc_xyb} gives a tool for subsequent elaborations.

Indeed, the above equations can be further developed using the Blatter-Geshkenbein-Larkin (BGL) scaling law:\cite{BGL,scaling2,blatterone} in the London approximation, a thermodynamic or intrinsic transport property $q$ of a uniaxially anisotropic superconductor depends on the applied field $B$ and angle $\theta$ only through the ratio $B/B_{c2}(\theta)$, so that $q(B,\theta)=s_q q^{iso}(B\epsilon(\theta))$, where $q^{iso}(B)$ is the field-dependent quantity in the equivalent (according to the scaling rules) isotropic superconductor, and $s_q$ is a scaling factor typically equal to 1, $\gamma$ or $\epsilon^{\pm1}(\theta)$. 

The current densities $J_c^{(x)}$ and $J_c^{(y_d)}$ are provided in Ref. \onlinecite{blatterone} for point-pinning in the single vortex and small-bundle pinning regime (the scaling theory does not describe the large bundle pinning regime, in the highest field and temperature regions). It is found\cite{blatterone} that they scale as:
\begin{subequations}
\label{eq:scalingJc}
\begin{align}
J_c^{(x)}(B,\theta)&=J_c^{iso}(B\epsilon(\theta))\\
  J_c^{(y_d)}(B,\theta)&=\epsilon(\theta)J_c^{iso}(B\epsilon(\theta))
\end{align}
\end{subequations}
where the full expression of $J_c^{iso}(B)$ is reported in Ref. \onlinecite{blatterone} for different pinning regimes.

It is worth stressing that the critical current $J_c^{iso}(B)$ (and the corresponding $k_p^{iso}(B)=c{\Phi_0 J_c^{iso}(B)}/{\xi^{iso}}$) for the equivalent isotropic superconductor is not angle-dependent because point pins only are considered, whereas extended pinning centers would introduce preferential directions and therefore angle-dependent quantities even in the isotropic superconductor.

Geometrically, the quantities $\xi^{(y_d)}(\theta)$ and $\xi^{(x)}(\theta)$ represent the maximum distances between the axis of the displaced vortex and the point pin. These distances are univocally defined since in the present model vortices are straight lines and the pinning centers are points with zero spatial extension. They scale as:\cite{blatterone}
\begin{subequations}
\label{eq:scalingxi}
\begin{align}
  \xi^{(y_d)}(\theta)&=\epsilon(\theta)\xi\\
  \xi^{(x)}(\theta)&=\xi 
\end{align}
\end{subequations}
where $\xi=\xi^{iso}$ denotes the in-plane coherence length. 

By using Eqs. \eqref{eq:scalingJc} and \eqref{eq:scalingxi} in Eq. \eqref{eq:kp_Jc_xy}, one can write:
\begin{subequations}
\label{eq:kp_Jc_xy2}
\begin{align}
\label{eq:kp_Jc_xy2a}
  k_p^{(x)}(B,\theta)&=s_{k_p^{(x)}} k_p^{iso}(B\epsilon(\theta)) \text{, } s_{k_p^{(x)}}={1}/{\epsilon(\theta)}\\
  k_p^{(y_d)}(B,\theta)&=s_{k_p^{(y_d)}} k_p^{iso}(B\epsilon(\theta)) \text{, } s_{k_p^{(y_d)}}=\epsilon(\theta)
\end{align}
\end{subequations}
It is worth stressing that, thanks to Eq. \eqref{eq:rhoC_i1}, it is 
\begin{equation}
\label{eq:kp11}
  k_{p,11}^\i(B, \theta)=k_p^{(x)}(B, \theta)
\end{equation}
hence the in-plane pinning constant $k_{p,11}^\i(\theta)$ obeys the scaling law given by Eq. \eqref{eq:kp_Jc_xy2a}.

\subsection{The Campbell regime tensors}

According to Eqs. \eqref{eq:rhoC_xya} and \eqref{eq:kp_Jc_xy2a}, one obtains:
\begin{equation}
\label{eq:rhoCscaling}
  \rho_{C,11}^\i(B,\theta)=s_{\rho_{C,11}} \rho_{C}^{iso}(B\epsilon(\theta)) \text{, } s_{\rho_{C,11}}=1
\end{equation}
On the other hand, $\rho_{C,33}^{(i)}$, using Eqs. \eqref{eq:rhoC_i}, \eqref{eq:rhoC_xy} and \eqref{eq:kp_Jc_xy2}, becomes:
\begin{align}
\label{eq:rhoC_i33} 
\nonumber
  \rho_{C,33}^\i(B,\theta)&=\frac{\rho_{C,11}^\i(B,\theta)\sin^2\theta}{\frac{k_{p}^{(y_d)}(B,\theta)}{k_p^{(x)}(B,\theta)}-\cos^2\theta}=\frac{\rho_{C,11}^\i(B,\theta)\sin^2\theta}{\epsilon^2(\theta)-\cos^2\theta}=\\
  &=\rho_{C,11}^\i(B/B_{c2}(\theta))\gamma^2
\end{align}
Therefore, similarly to $\dbar{\rho}_\ff^\i$ of Eq. \eqref{eq:rhoff_tensor2}, one can write: 
\begin{equation}
\label{eq:rhoC_tensor}
  \dbar{\rho}_{C}^\i(B,\theta)=\rho_{C,11}^\i(B/B_{c2}(\theta))\dbar{M}
\end{equation}
This is an important result of this paper: for point pinning, the anisotropy of the pinning tensors and the flux flow tensors is the same and is completely described by the mass anisotropy tensor. 
This result is very important also in the study of regimes where both losses and pinning phenomena are equally relevant, placing themselves in a intermediate situation between purely dissipative flux flow and purely reactive Campbell regimes. These regimes, occurring at higher frequencies including the microwave range, will be studied in a future work.\cite{Pompeo_Part2}

Lastly, due to Eq. \eqref{eq:rhoC_i33}, the intrinsic pinning constant tensor also satisfies the scaling law:
\begin{equation}
\label{eq:kptensor_Bdep}
\dbar{k}_{p}^\i(B,\theta)=k_{p,11}^\i(B/B_{c2}(\theta))\dbar{M}^{-1}\\
\end{equation}
while the measured pinning constant tensor does not: 
\begin{equation}
\dbar{k}_{p}(B,\theta)=-\xma{B}\dbar{k}_{p}^\i\xma{B}=k_{p,11}^\i(B/B_{c2}(\theta))\dbar{\mathcal{M}}(\theta,\phi)
\end{equation}
because of the additional contribution given by $\dbar{\mathcal{M}}(\theta,\phi)$ (Eq. \eqref{eq:MB}).

\section{Application to experiment: a.c. linear susceptibility measurements analysis}
\label{sec:exp_Campbell}

In this Section I provide an example showing the additional information which can be gained using the results of the present work by performing an analysis of experimental data.
Various techniques can be used to explore the Campbell regime, including inductance measurements,\cite{inductive} a.c. linear susceptibility measurements \cite{suscept} and the vibrating reed technique.\cite{vibratingreed} The pinning constant can be also determined through microwave measurements,\cite{omegap_kp_HTS,more_kp} but they typically go beyond the pure Campbell regime and, as such, they are beyond the scope of this work and postponed to a future study.\cite{Pompeo_Part2}
In the following, I focus on a.c. linear susceptibility measurements on YBa$_2$Cu$_3$O$_{7-\delta}$ samples in the mixed state, performed by varying both the direction and the intensity of the applied d.c. magnetic field. The source of the experimental data is Ref. \onlinecite{Pasquini}, where full details about the experiment can be found. The measured squared real part of the penetration depth $\lambda^2_R=\lambda^2_L+\lambda_C^2$ ($\lambda_L$ is the London penetration depth) is related to the squared Campbell penetration depth $\lambda_C^2$ which, given the geometry of the experiment, is $\lambda_{C,11}\propto k_{p,11}^{-1}$.
The examined sample are two twinned YBa$_2$Cu$_3$O$_{7-\delta}$ single crystals, one irradiated at +30$^\circ$ with respect the $c$-axis in order to create columnar defects capable of reinforcing the pinning properties and one, virgin, used as reference.
As a consequence, four sources of pinning are expected: point pins and three types of correlated pinning centers, namely the twins along the $c$-axis direction, the $a$-$b$ planes and the columnar defects at +30$^\circ$ (in the irradiated sample only). It was concluded that pinning was stronger with the field aligned with columnar defects and, from a qualitative analysis, that an enhancement of pinning existed even far from the track direction. I show in the following that such findings can be put on solid quantitative grounds by exploiting the present model. In particular, I will use the model developed in the previous Section to remove the anisotropic response due to the material anisotropy and to point pins, which can give rise to a significant angle-dependent contribution. The latter can in principle mask the angle-dependent contributions arising from correlated defects.

Here I consider the data taken from Fig. 5 of Ref. \onlinecite{Pasquini} for the irradiated sample only, where $\lambda^2_R(B)$ measured at various angles $\theta$ at T=$90.5\;$K is reported. In order to extract $k_{p,11}(B,\theta)$, I extrapolate $\lambda^2_R(B\rightarrow 0)$, take it as an evaluation of $\lambda^2_L$ and, by neglecting pair-breaking effects, estimate $\lambda_C^2(B)=\lambda^2_R(B)-\lambda^2_R(B\rightarrow 0)$. Then the pinning constant is obtained, using Eq. \eqref{eq:rhoC} with $B=\mu_0H$ and normalizing with $k_{p,11}^{max}$. The normalized pinning constant, $k_{p,11}/k_{p,11}^{max}$, is then free from geometrical factors.

\begin{figure}[ht]
\centerline{\includegraphics[width=8cm]{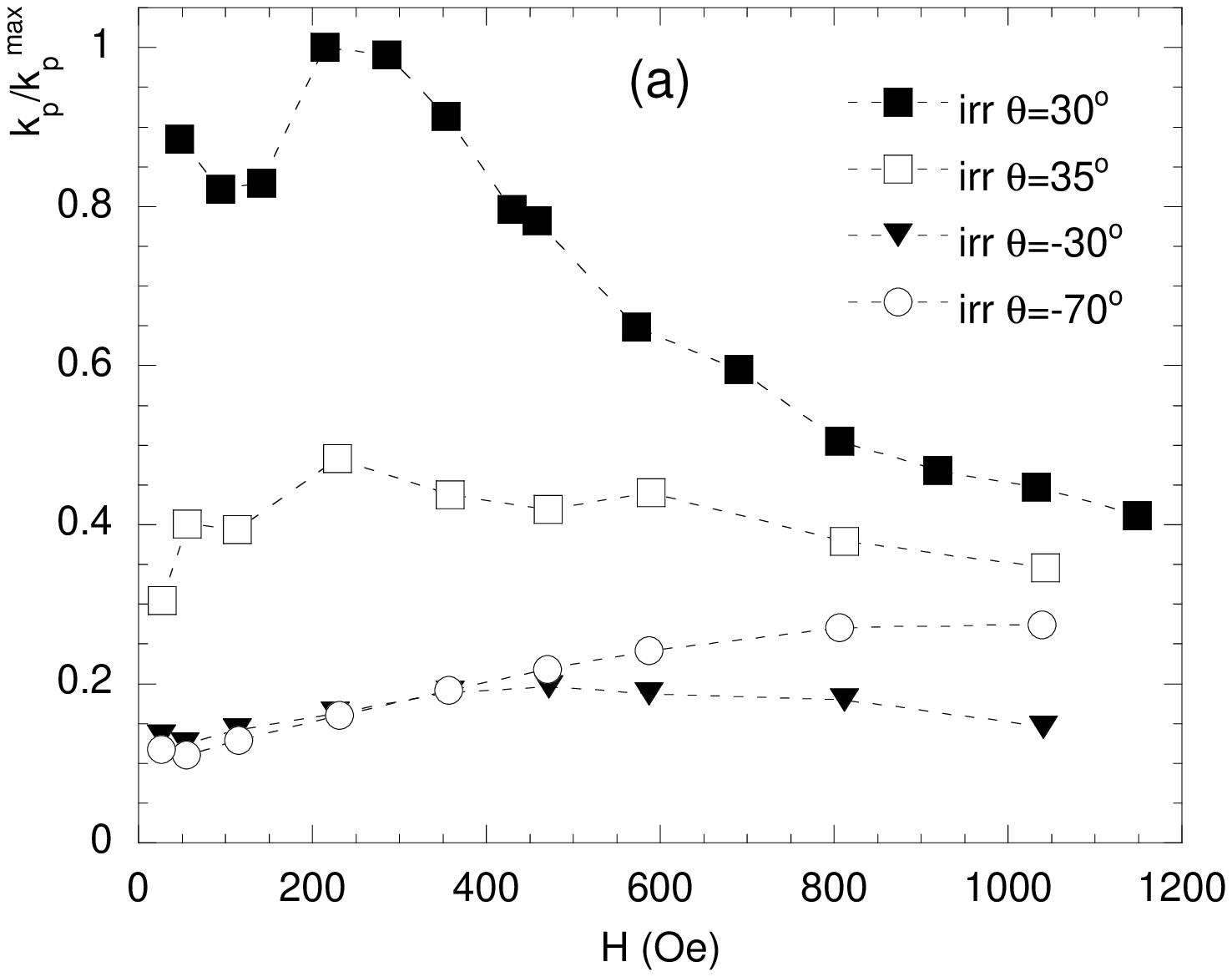}}
\centerline{\includegraphics[width=8cm]{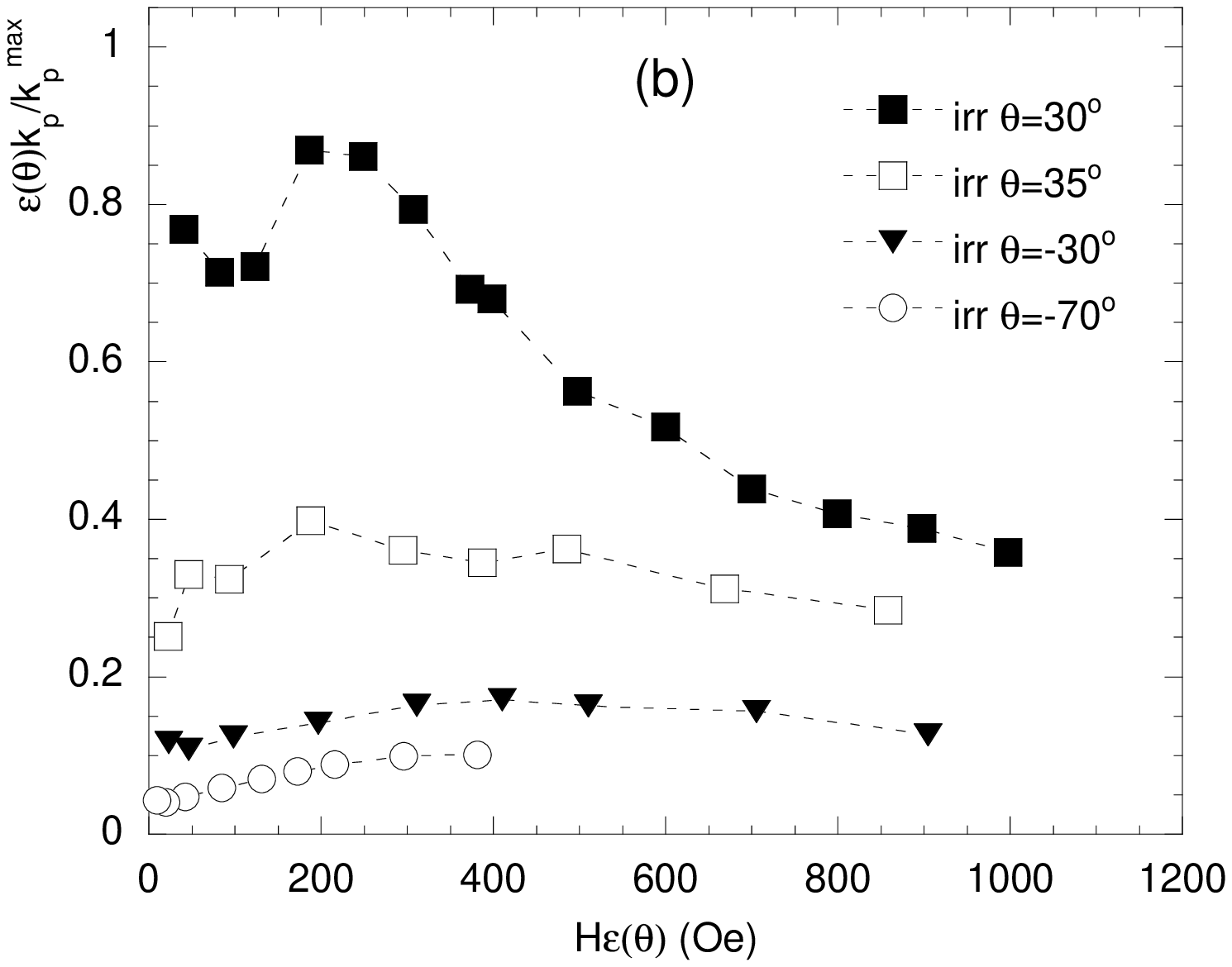}}
  \caption{Normalized pinning constant, as is (a) and scaled (b), as a function of d.c. applied field at selected angles. (a): bare quantities; (b): scaled quantities. Data digitized from Ref. \onlinecite{Pasquini} as reported in the text.}
\label{fig:labusch}
\end{figure}

The result is reported in Fig. \ref{fig:labusch}a. The pinning constant in the irradiated sample with $\widehat{B}$ along the columnar defects ($\theta=+30^\circ$) is larger than along the other directions. Moreover, by examining the curves with $\theta=-30^\circ$ and $\theta=-70^\circ$, it was noted\cite{Pasquini} that, despite being very similar at low fields, the two curves depart near half the matching field $B_\Phi\approx350\;$Oe. Thus, it was inferred\cite{Pasquini} that the defects still determined an appreciable contribution to pinning even for the field tilted at $\theta=-30^\circ$, 60$^\circ$ far from the tracks directions. 
Actually, this observation can be further substantiated by resorting to the result of Section \ref{sec:campbell}, namely Eqs. \eqref{eq:kp_Jc_xy2a} and \eqref{eq:kp11}: were only point pinning present, the in-plane pinning constant at different fields and angles should scale so that by plotting $\epsilon(\theta) k_{p,11}$ vs $B\epsilon(\theta)$ all the curves should overlap. Any possible residual difference between the scaled curves should then be ascribed to the effects of correlated defects only, since the scaling removes the background effect of point pinning centers. The result of such scaling is reported in Fig. \ref{fig:labusch}b, where $\gamma=7$ has been used (taking\cite{poole} $\gamma=5\div8$ does not change the result). It can be seen that, even by having removed the effect of point pins, along the tracks directions pinning remains the largest. On the 
other hand, the absolute values for $\theta=-30^\circ$ are now larger than those for  $\theta=-70^\circ$. Since at these 
angles the contribution of twins and $a$-$b$ planes can be neglected (see also Fig. 2 in Ref. \onlinecite{Pasquini}), one can infer that at $\theta=-30^\circ$ the artificial defects still reinforce pinning, yielding a higher pinning constant than the one measured at $\theta=-70^\circ$. This result further substantiates, on quantitative grounds, the observation done on qualitative grounds, i.e. based on the field dependence, reported in the original work.\cite{Pasquini}

\section{Summary}
\label{sec:summary}
The electrodynamics model for transport measurements in the mixed state in uniaxial anisotropic superconductors has been discussed in the very different free flux flow and pinned (with point pins) Campbell regimes. 
Vortex parameters, like the viscous drag, the vortex mobility and pinning constant, have been derived in tensor form for arbitrary field orientations. It has been shown that the tensors describing point pinning share the same structure of the flux flow tensors, and that the measured quantities, differently from the corresponding intrinsic quantities, in general do not satisfy the angular scaling laws.
It has also been shown that the experiments do not generally yield the intrinsic values of the anisotropic viscosity and pinning constant. Explicit expressions relating measured and intrinsic quantities have been given. 
An example of data analysis based on the results obtained in the Campbell regime has been provided. The results here obtained will prove necessary to investigate the angular dependence of the complex a.c. resistivity.\cite{Pompeo_Part2}

\acknowledgments
This work has been supported by Regione Lazio and partially supported by the Italian FIRB project ``SURE:ARTYST'' and by EURATOM.

The author warmly acknowledges the helpful discussion with prof. E. Silva.

\appendix
\section{Tensor Algebra Notes}
\subsection{Cross product}
\label{sec:crossproduct}
The cross product $\vec{a}\times\vec{b}$ between a generic vector $\vec{a}$ and another vector $\vec{b}$ can be also computed as the application of a tensor $\xma{a}$ to $\vec{b}$:
\begin{equation}
\nonumber
\vec{a}\times\vec{b}=\xma{a}\vec{b}
\end{equation}
where $\xma{a}$ is defined starting from $\vec{a}\;$:\cite{tensors2}
\begin{equation}
\label{eq:crossproduct}
\xma{a}:=
\begin{pmatrix}
  0 & -a_{z} & a_{y} \\
  a_{z} & 0 & -a_{x} \\
  -a_{y} & a_{x} & 0
\end{pmatrix}
\nonumber
\end{equation}

\subsection{Factoring out tensors}
\label{sec:factoringtensors}
Here the following property is demonstrated:

\begin{equation}
\label{eq:factoringtensors}
\left(\dbar{A}\vec{b}\right)\times\left(\dbar{A}\vec{c}\right)=\left|\dbar{A}\right|\dbar{A}^{-1}\left(\vec{b}\times\vec{c}\right)\\ 
\end{equation}
which holds $\forall\;\vec{b},\vec{c}$ and with $\dbar{A}=\diag(a_{11},a_{22},a_{33})$ an invertible {\em diagonal} tensor.

Since it is easy to verify that
\begin{equation}
\begin{gathered}
\nonumber
\dbar{A}\left[\left(\dbar{A}\vec{b}\right)\times\left(\dbar{A}\vec{c}\right)\right]=
\begin{vmatrix}
  a_{11}\widehat{x} & a_{22}\widehat{y} & a_{33}\widehat{z} \\
  a_{11}b_x & a_{22}b_y & a_{33}b_z \\
  a_{11}c_x & a_{22}c_y & a_{33}c_z
\end{vmatrix}=\\ \nonumber
=a_{11}a_{22}a_{33}
\begin{vmatrix}
  \widehat{x} & \widehat{y} & \widehat{z} \\
  b_x & b_y & b_z \\
  c_x & c_y & c_z
\end{vmatrix}
=\left|\dbar{A}\right|\left(\vec{b}\times\vec{c}\right)
\end{gathered}
\end{equation}
the thesis straightforwardly follows.

\section{Measurable flux flow resistivity tensor according to HHT model}
\label{sec:HHT_rhoff}

In Refs. \onlinecite{HHTelectrodyn,HHTelectrodyn2}, HHT provide an elegant and compact way to derive, from the intrinsic flux flow resistivity reviewed in Sec. \ref{sec:rhoffi_tensor}, the experimentally measured (and thus effective) resistivities, which relate the actual {\em externally applied} current density $\vec{J}$ to the electric field induced by vortex motion, considered in terms of its components along the directions both parallel and normal to the current. 
Here this model is reviewed and cast in tensor form, showing that it provides the same expression as Eq. \eqref{eq:rhoff_uni} obtained from the force Equation. 

By isolating $\vec{J}_T$ in Eq. \eqref{eq:JT_def2}, substituting it in Eq. \eqref{eq:rhoff_def} and considering that $\vec{J}_S=J_S\widehat{B}$, the following identity holds:
\begin{equation}
\label{eq:JS0}
\dbar{\rho}_\ff^\i\vec{J}_T=\dbar{\rho}_\ff^\i\vec{J}-\dbar{\rho}_\ff^\i\widehat{B}J_S
\end{equation}
and by taking the dot product by $\widehat{B}$, considering Eq. \eqref{eq:JT_def1}, one obtains:
\begin{equation}
\label{eq:JS}
\vec{J}_S=\frac{\left(\dbar{\rho}_\ff^\i\vec{J}\right)\cdot\widehat{B}}{\left(\dbar{\rho}_\ff^\i\widehat{B}\right)\cdot\widehat{B}}\widehat{B}
\end{equation}
It is worth noting that $\vec{J}_s$ and $\vec{J}_T$ are different from the components of $\vec{J}$ respectively parallel ($\vec{J}_{\parallelslant}$) and perpendicular ($\vec{J}_{\bot}$) to $\widehat{B}$: in general $\vec{J}_T$ has itself a component parallel to $\widehat{B}$. Only when the superconductor is isotropic one has  $\vec{J}_s=\vec{J}_{\parallelslant}$ and $\vec{J}_T=\vec{J}_{\perp}$, as it can be checked considering Eqs. \eqref{eq:JT_def1} and \eqref{eq:JS}. 

Substituting Eq. \eqref{eq:JS} together with Eq. \eqref{eq:JT_def2} into Eq. \eqref{eq:rhoff_def} yields, with some tensor algebra:
\begin{equation}
\nonumber
\vec{E}=\frac{1}{\left(\dbar{\rho}_\ff^\i\widehat{B}\right)\cdot\widehat{B}}\left\{\widehat{B}\times\left[\left(\dbar{\rho}_\ff^\i\vec{J}\right)\times\left(\dbar{\rho}_\ff^\i\widehat{B}\right) \right]\right\}
\end{equation}
This equation allows to determine $\dbar{\rho}_{\ff}$ defined in Eq. \eqref{eq:rhoeff_def}. Since  $\dbar{\rho}_\ff^\i$ is diagonal the above Equation can be rewritten by exploiting property  \eqref{eq:factoringtensors}:
\begin{eqnarray}
\label{eq:Erhodiag}
\vec{E}&=&\frac{1}{\left(\dbar{\rho}_\ff^\i\widehat{B}\right)\cdot\widehat{B}}\left\{\widehat{B}\times\left[\left(\left|\dbar{\rho}_\ff^\i\right|\dbar{\rho}_\ff^{\i-1}\right)\left(\vec{J}\times\widehat{B}\right) \right]\right\}= \nonumber \\
&=& -\frac{\left|\dbar{\rho}_\ff^\i\right|}{\left(\dbar{\rho}_\ff^\i\widehat{B}\right)\cdot\widehat{B}}\left(\xma{B}\dbar{\rho}_\ff^{\i-1}\xma{B}\right)\vec{J}
\end{eqnarray}
By comparing Eq. \eqref{eq:rhoeff_def} and Eq. \eqref{eq:Erhodiag}, one can write down explicitly $\dbar{\rho}_{\ff}$:
\begin{align}
\label{eq:rhopff}
 \nonumber
\dbar{\rho}_{\ff}(\theta,\phi)&=-\xma{B}\left(\frac{\left|\dbar{\rho}_\ff^\i\right|\dbar{\rho}_\ff^{\i-1}}{\left(\dbar{\rho}_\ff^\i\widehat{B}\right)\cdot\widehat{B}}\right)\xma{B}=\\ \nonumber
&=-\xma{B}\left(\frac{\rho_{\ff,11}^\i\left|\dbar{M}\right|}{\left(\dbar{M}\widehat{B}\right)\cdot\widehat{B}}\dbar{M}^{-1}\right)\xma{B}=\\ \nonumber
&=-\xma{B}\left(\frac{\rho_{\ff,11}^\i}{\epsilon^2}\dbar{M}^{-1}\right)\xma{B}=\\
&=\frac{\rho_{\ff,11}^\i(\theta,\phi)}{\epsilon^2(\theta,\phi)}\dbar{\mathcal{M}}(\theta,\phi)
\end{align}
which is equal to Eq. \eqref{eq:rhoff_uni}.

\section{The out-of-plane intrinsic flux flow resistivity cannot be measured with a single setup}
\label{sec:tensorelementsderivation}

Without loss of generality I consider the static field $\vec{B}$ oriented according to the angles $(\theta,\phi)$ and a current density $\vec{J}=J\widehat{J}\in x$-$z$ plane, with unit vector $\widehat{J}=(\sin\alpha,0,\cos\alpha)$ ($\alpha$ is the angle between $\vec{J}$ and the $z$ axis). Despite the choice $\vec{J}\in x$-$z$ plane, the relative orientation of $\vec{B}$ and $\vec{J}$ is completely free, as long as $\alpha$ and $\phi$ are free angles.
In this configuration, the measured flux flow resistivity along $\vec{J}$ is, according to Eq. \eqref{eq:rhoJ}:
\begin{align}
\label{eq:rhoJalpha}
\nonumber
\rho^{\widehat{J}}&(\theta,\phi,\alpha)=\frac{\rho_{\ff,11}^\i(\theta)}{\rho_{\ff,11}^\i(\theta)\sin^2\theta+\rho_{\ff,33}^\i(\theta)\cos^2\theta}\cdot \\ \nonumber
&(\rho_{\ff,11}^\i(\theta)\sin^2\phi\sin^2\theta\sin^2\alpha+\rho_{\ff,33}^\i(\theta)(\cos^2\theta\sin^2\alpha+\\  \nonumber
&-2\cos\phi\cos\theta\sin\theta\cos\alpha\sin\alpha+\sin^2\theta\cos^2\alpha))=\\ \nonumber
&=\frac{\rho_{\ff,11}^\i(\theta)}{\rho_{\ff,11}^\i(\theta)\sin^2\theta+\rho_{\ff,33}^\i(\theta)\cos^2\theta}\cdot \\ 
&[a_1(\phi,\theta,\alpha)\rho_{\ff,11}^\i(\theta)+a_3(\phi,\theta,\alpha)\rho_{\ff,33}^\i(\theta)]
\end{align}
where $a_1(\phi,\theta,\alpha)$ and $a_3(\phi,\theta,\alpha)$ represent in compact way the combination of trigonometric functions appearing in the second equality. 
In order to obtain $\rho_{\ff,33}^\i(\theta)$ from the measured $\rho^{\widehat{J}}(\theta,\phi,\alpha)$ without being forced to know also $\rho_{\ff,11}^\i(\theta)$, in the last equality $\rho_{\ff,11}^\i(\theta)$ must disappear for each $\theta$ with the freedom to choose properly $\phi$ and $\alpha$. This is impossible since the term in square parentheses should be equal to a rational function of $\rho_{\ff,11}^\i$ and $\rho_{\ff,33}^\i$.

On the other hand, it can be easily checked that it is possible to obtain $\rho_{\ff,11}^\i$ only: it is sufficient to choose $\phi$ and $\alpha$ (e.g. maximum Lorentz force configuration: $\phi=0$ and $\alpha=\pi/2$) so that the term in square parentheses simplifies with the denominator of the fraction.

\end{document}